\documentclass[11pt]{llncs}

\usepackage{amsmath,latexsym}
\usepackage{amsfonts}
\usepackage{amssymb}
\usepackage{amssymb}
\usepackage{graphicx}
\usepackage{epstopdf}
\usepackage{subfigure}
\usepackage{setspace}
\usepackage{color}
\usepackage{setspace}
\usepackage{algpseudocode}
\usepackage{algorithm}
\newtheorem{thrm}{Theorem}
\newtheorem{lemm}[thrm]{Lemma}

\newtheorem{defn}[thrm]{Definition}
\newtheorem{obs}[thrm]{Observation}

\def\qedbox#1#2{\vbox{\hrule height.2pt
  \hbox{\vrule width.2pt height#2pt \kern#1pt \vrule width.2pt}
  \hrule height.2pt}}

\def\s#1{\mbox{\boldmath $#1$}}

\def\+{\!+\!}
\def\-{\!-\!}

\def\m{\!-\!}

\def\itbf#1{\textit{\textbf{#1}}}

\def\bproc{{\bf procedure\ }}

\def\bfor{{\bf for\ }}

\def\bto{{\bf to\ }}

\def\bwhile{{\bf while\ }}

\def\bdo{{\bf do\ }}
\def\bif{{\bf if\ }}
\def\bthen{{\bf then\ }}
\def\belse{{\bf else\ }}

\def\la{\leftarrow}

\def\qq{\qquad}
\def\com#1{{\bf $\triangleright$}\hspace{6pt}{\sl #1}}

\def\pref(#1,#2){$#1$ is a prefix of $#2$}
\def\suff(#1,#2){$#1$ is a suffix of $#2$}

\def\reg(#1,#2){$#2$ is $#1$-regular}
\def\notreg(#1,#2){$#2$ is not $#1$-regular}

\def\UPDATE\_F{\tt{UPDATE\_F}}

\def\mec{\tt{mec}}
\def\MEC{\tt{MEC}}
\def\CMEC{\tt{CMEC}}
\def\B{\tt{B}}
\def\Q'{\tt{Q'}}

\def\MNC{\tt{MNC}}
\def\PR{\tt{PR}}
\def\CPR{\tt{CPR}}
\def\POS{\tt{POS}}
\def\LEN{\tt{LEN}}

\newif\ifShow
\Showfalse

\algnewcommand{\LineComment}[1]{\State \(\triangleright\) \normalfont{\sl #1}}
\algtext*{EndWhile}
\algtext*{EndIf}
\algtext*{EndFor}
\algtext*{EndFor}
\algtext*{EndProcedure}

\begin{document}

\pagestyle{headings}
\title{Enhanced Covers of Regular \& Indeterminate\\ Strings using Prefix Tables}

\author{
 Ali Alatabbi\inst{1}
 \and
 A.\ S.\ Sohidull Islam\inst{2}
 \and
 M.\ Sohel Rahman\thanks{Supported in part by a Commonwealth Academic Fellowship and a ACU Titular Fellowship. Currently on a sabbatical leave from BUET.}\inst{1,3}
 \and  \\
 Jamie Simpson\inst{4,5}
 \and
 W.\ F.\ Smyth\thanks{Supported in part by a grant from the Natural Sciences \& Engineering
 Research Council (NSERC) of Canada.}\inst{2,5}}
 
 \institute{Department of Informatics, King's College London\\
 \email{ali.alatabbi@kcl.ac.uk}
 \and Algorithms Research Group, Department of Computing \& Software\\
 McMaster University\\
 \email{sohansayed@gmail.com,smyth@mcmaster.ca}
 \and Department of Computer Science \& Engineering\\
 Bangladesh University of Engineering \& Technology\\
 \email{msrahman@cse.buet.ac.bd}
 \and
 Department of Mathematics and Statistics, Curtin University of Technology
 \email{Jamie.Simpson@curtin.edu.au}
 \and School of Engineering \& Information Technology\\
 Murdoch University
 }

\maketitle

\begin{abstract}
A \itbf{cover} of a string $\s{x} = \s{x}[1..n]$
is a proper substring \s{u} of \s{x} such that
\s{x} can be constructed from possibly overlapping instances of \s{u}.
A recent paper \cite{FIKPPST13} relaxes this definition --- an \itbf{enhanced cover} \s{u} of \s{x}
is a border of \s{x} (that is, a proper prefix that is also a suffix)
that covers a {\it maximum} number of positions in \s{x}
(not necessarily all) --- and
proposes efficient algorithms for the computation of enhanced covers.
These algorithms depend on the prior computation of the \itbf{border array} $\s{\beta}[1..n]$,
where $\s{\beta}[i]$ is the length of the longest border of $\s{x}[1..i]$,
$1 \le i \le n$.
In this paper, we first show how to compute enhanced covers using instead the \itbf{prefix table}:
an array $\s{\pi}[1..n]$ such that $\s{\pi}[i]$
is the length of the longest substring of \s{x} beginning at position $i$
that matches a prefix of \s{x}.
Unlike the border array, the prefix table is robust: its properties
hold also for \itbf{indeterminate strings} --- that is,
strings defined on {\it subsets} of the alphabet $\Sigma$
rather than individual elements of $\Sigma$.
Thus, our algorithms, in addition to being faster in practice and more space-efficient than those of \cite{FIKPPST13},
allow us to easily extend the computation of enhanced covers
to indeterminate strings.
Both for regular and indeterminate strings,
our algorithms execute in expected linear time.
Along the way we establish an important theoretical result:
that the expected maximum length of any border of any prefix of a regular string \s{x}
is approximately 1.64 for binary alphabets, less for larger ones.
\end{abstract}

\section{Introduction}
\label{sect-intro}
The concept of {\it periodicity} is fundamental to combinatorics on words
and related algorithms: it is difficult to imagine a research contribution
that does not somehow involve periods of strings.
But periodicity alone may not be the best descriptor of a string;
for example, $\s{x} = abaababab$, a string of length $n = 9$, has period $7$
and corresponding \itbf{generator}\footnote{Notation and terminology generally follow \cite{S03}.}
$abaabab$, but it could well be more interesting that every position but one in \s{x}
lies within an occurrence of $ab$.
In 1990 Apostolico \& Ehrenfeucht \cite{AE90} introduced the idea
of quasiperiodicity: a \itbf{quasiperiod} or \itbf{cover} of a string \s{x}
is a proper substring \s{u} of \s{x} such that any position in \s{x}
is contained in an occurrence of \s{u};
\s{u} is then said to \itbf{cover} \s{x},
which is said to be \itbf{quasiperiodic}.
Thus, for example, $\s{u} = aba$ is a cover of $\s{x} = ababaaba$.
Several linear-time algorithms were proposed for the computation of covers
\cite{AFI91,B92,MS94,MS95}, culminating in an algorithm \cite{LS02}
to compute the \itbf{cover array} \s{\gamma},
where $\s{\gamma}[i]$ gives the length $j$ of the longest cover of $\s{x}[1..i]$.
Since the longest cover of $\s{x}[1..j]$ is also a cover of $\s{x}[1..i]$,
\s{\gamma} implicitly specifies all the covers of every prefix of \s{x}.
A recent paper \cite{ARS14a} extends the computation of \s{\gamma}
to ``indeterminate strings'' (see below for definition).

Even though the cover of a string can provide useful information,
quasiperiodic strings are on the other hand infrequent among strings in general.
Another approach to string covering was therefore proposed in \cite{IS98}:
a set $U_k = U_k(\s{x})$ of strings, each of length $k$, is said to be a \itbf{minimum $k$-cover}
of \s{x} if every position in \s{x} lies within some occurrence of an element of $U_k$,
and no smaller set of $k$-strings has this property.
Thus $U_2(abaababab) = U_2(ababaaba) = \{ab,ba\}$.
In \cite{CIMSY05} the computation of $U_k$ was shown to be NP-complete,
though an approximate polynomial-time algorithm was presented in \cite{IMS11}.

Recall that a \itbf{border} of \s{x} is a possibly empty proper prefix of \s{x}
that is also a suffix: every nonempty string has a border of length zero.
Recently the promising idea of an \itbf{enhanced cover} was introduced \cite{FIKPPST13}; that is,
a border \s{u} of $\s{x} = \s{x}[1..n]$ that covers a maximum number $m \le n$ of positions in \s{x}.
Then the \itbf{minimum enhanced cover} $\mec(\s{x})$ is the {\it shortest} border \s{u}
that covers $m$ positions,
and \cite{FIKPPST13} presented an algorithm to compute $\mec(\s{x})$
in $\Theta(n)$ time.
Thus for $\s{x} = abaababab$, $\mec(\s{x}) = ab$.
Further, on the analogy of the cover array defined above, the authors
proposed the \itbf{minimum enhanced cover array} $\MEC_{\s{x}}$ ---
for every $i \in 1..n$, $\MEC_{\s{x}}[i] = |\mec(\s{x}[1..i])|$,
the length of the minimum enhanced cover of $\s{x}[1..i]$ ---
and showed how to compute it in $\mathcal{O}(n\log n)$ time.
In this paper we introduce in addition the $\CMEC$ array,
where $\s{\CMEC}[i]$ specifies the number of positions in \s{x}
covered by the border of length $\s{\MEC}[i]$.
Thus, for example, $\MEC_{abaababab} = 001123232$
and $\s{\CMEC}_{abaababab} = 002346688$.

In order to compute $\MEC_{\s{x}}$, the authors of \cite{FIKPPST13} made use of a variant
of the \itbf{border array} --- that is, an integer array $\s{\beta}[1..n]$
in which for every $i \in 1..n$, $\s{\beta}[i]$ is the length of the
longest border of $\s{x}[1..i]$.
In this paper we adopt a different approach to the computation of $\MEC_{\s{x}}$,
using, instead of a border array, the \itbf{prefix table} $\s{\pi} = \s{\pi}[1..n]$,
where for every $i \in 1..n$, $\s{\pi}[i]$ is the length of the longest substring at
position $i$ of \s{x} that equals a prefix of \s{x}.
It has long been folklore that \s{\beta} and \s{\pi} are ``equivalent'',
but it has only recently been made explicit \cite{BKS13} that
each can be computed from the other in linear time.
However, this equivalence holds only for \itbf{regular} strings \s{x} in which
each entry $\s{x}[i]$ is constrained to be a single element of the underlying alphabet $\Sigma$.

We say that a letter $\lambda$ is \itbf{indeterminate} if it is
any nonempty subset of $\Sigma$, and thus a string \s{x} is said to be
\itbf{indeterminate} if some constituent letter $\s{x}[i]$ is indeterminate.
The idea of an indeterminate string was first introduced in \cite{FP74} --- with letters
constrained to be either regular (single elements of $\Sigma$) or $\Sigma$ itself ---
and the properties of these strings have been much studied by Blanchet-Sadri \cite{BS08} and her collaborators
as ``partial words'' or ``strings with holes''.
Indeterminate strings can model DNA sequences on $\Sigma = \{A,C,G,T\}$
when ambiguities arise in determining individual nucleotides (letters).

Two indeterminate letters $\lambda$ and $\mu$ are said to \itbf{match}
(written $\lambda \approx \mu$)
whenever $\lambda \cap \mu \ne \emptyset$,
a relation that is in general nontransitive \cite{HS03,SW09}:
$a \approx \{a,b\}$ and $\{a,b\} \approx b$, but $a \not\approx b$.
An important consequence of this nontransitivity is that the border array
no longer correctly describes {\it all} of the borders of \s{x}:
it is no longer necessarily true, as for regular strings, that
if \s{u} is the longest border of \s{v}, in turn the longest
border of \s{x}, then \s{u} is a border of \s{x}.
On the other hand, the prefix array retains all its properties for indeterminate strings \s{x}
and, in particular, correctly identifies all the borders of every prefix of \s{x} \cite{BKS13}.
\cite{SW08} describes
algorithms to compute the prefix table of an indeterminate string;
conversely, \cite{CRSW13} proves that there exists an indeterminate string
corresponding to every feasible prefix table,
while \cite{ARS14} describes an algorithm to compute the
lexicographically least indeterminate string determined by
any given feasible prefix table.
There is thus a many-many relationship between
the set of all indeterminate strings and the set of all prefix tables.
Consequently, computing $\MEC_{\s{x}}$ (or simply $\MEC$ when there is no ambiguity)
from the prefix table $\s{\pi} = \s{\pi}_{\s{x}}$
rather than from a variant of the border array
allows us to extend the computation to indeterminate strings.

In Section~\ref{sect-method} we outline the basic methodology
and data structures used to compute the minimum enhanced cover array from the prefix table,
while illustrating the ideas with an example.
Then Section~\ref{sect-analysis} provides
a proof of the algorithm's correctness,
as well as an analysis of its complexity, both worst and average case.
In Section~\ref{sect-compare} we discuss the practical application of
our algorithms, in terms of time and space requirements,
and compare our prefix-based implementation with the border-based implementation
of \cite{FIKPPST13}.
Section~\ref{sect-extend} extends the enhanced cover array algorithm
to indeteterminate strings (for rooted covers) and outlines
various other extensions,
particularly to generalizations of $\MEC$s.

\section{Methodology}
\label{sect-method}

In this section we describe the computation of $\MEC_{\s{x}}$,
the enhanced cover array of \s{x}, based on the prefix array $\s{\pi}$.
Since every minimum enhanced cover of \s{x} is also a border of \s{x},
we are initially interested in the covers of prefixes of \s{x}.
For this purpose we need arrays whose size is $\B$,
the maximum length of any border of any prefix of \s{x}.
Noting that $\B$ must be the maximum entry in the prefix array $\s{\pi}$, we write
$\B = \max_{2\le i \le n}\s{\pi}[i]$.
\begin{defn}
\label{def-MNC}
In the \itbf{maximum no cover} array $\s{\MNC} = \s{\MNC}[1..\B]$,
for every $q \in 1..\B$, $\s{\MNC}[q] = q'$,
where $q'$ is the maximum integer in $1..q$ such that
the prefix $\s{x}[1..q']$ has no cover --- that is,
such that $\s{\gamma}[q'] = 0$.
\end{defn}

As shown in Figure~\ref{fig-MNC},
once $\B$ is computed in $\Theta(n)$ time from the prefix array $\s{\pi}$,
$\s{\MNC}$ can be easily
computed in $\Theta(\B)$ time using the cover array
$\s{\gamma}[1..\B]$ of $\s{x}[1..\B]$.
Note that the entries in \s{\MNC} are monotone nondecreasing
with $1 \le \s{\MNC}[q] \le q$ for every $q \in 1..\B$.
The following is fundamental to the execution of our main algorithm:

\begin{obs}
\label{obs-mnc}
If a prefix $\s{v} = \s{x}[1..q]$ of \s{x} has a cover \s{u},
then $\s{v} \ne \mec(\s{x})$
(since $|\s{u}| < q$
and \s{u} covers every position covered by \s{v}).
\end{obs}

\begin{figure}[ht]
{\leftskip=0.80in\obeylines\sfcode`;=3000
\bproc Compute\_MNC$(n,\s{\pi};\ \B,\s{\gamma},\s{\MNC})$
$\B \la \s{\pi}[2]$
\bfor $i \la 3$ \bto $n$ \bdo
\qq $\B \la \max(\B,\s{\pi}[i])$
\com{Compute $\s{\gamma}[1..\B]$ of $\s{x}[1..\B]$ using}
\com{the algorithm Compute\_PCR of \cite{ARS14a}.}
Compute\_PCR$(\B,\s{\pi};\ \s{\gamma})$
\com{Note that \s{\MNC} can overwrite \s{\gamma}.}
\bfor $q \la 1$ \bto $\B$ \bdo
\qq \bif $\s{\gamma}[q] = 0$ \bthen $\s{\MNC}[q] \la q$
\qq \belse $\s{\MNC}[q] \la \s{\MNC}[q\m 1]$
}
\caption{Computing $\s{\MNC}$ from the prefix array $\s{\pi}[1..n]$ and the cover array $\s{\gamma}[1..B]$.}
\label{fig-MNC}
\end{figure}

Thus $\MNC[q]$ specifies an upper bound $q' \in 1..q$
on the length of a minimum enhanced cover of \s{x}.
Two other arrays are required for the computation,
both of length $\B$:
\begin{defn}
\label{def-PR/CPR}
For every $q \in 1..\B$:
\begin{itemize}
\item[$\bullet$]
$\PR[q]$ is the rightmost position in \s{x} at which
the prefix $\s{x}[1..q]$ occurs;
\item[$\bullet$]
$\CPR[q]$ is the number of positions in \s{x} covered
by occurrences of $\s{x}[1..q]$.
\end{itemize}
\end{defn}

Here is an example of the arrays introduced thus far:
\begin{equation*}
\begin{array}{rcccccccccc}
\scriptstyle 1 & \scriptstyle 2 & \scriptstyle 3 & \scriptstyle 4 & \scriptstyle 5 & \scriptstyle 6 & \scriptstyle 7 & \scriptstyle 8 & \scriptstyle 9 & \scriptstyle 10 \\
\s{x} = a & b & a & b & a & a & b & a & b & a \\
\s{\pi} = 10 & 0 & 3 & 0 & 1 & 5 & 0 & 3 & 0 & 1 \\
\s{\gamma} = 0 & 0 & 0 & 2 & 3 \\
\s{\MNC} = 1 & 2 & 3 & 3 & 3 \\
\s{\PR} = 10 & 8 & 8 & 6 & 6 \\
\s{\CPR} = 6 & 8 & 10 & 8 & 10 \\
\s{\MEC} = 0 & 0 & 1 & 2 & 3 & 1 & 2 & 3 & 2 & 3 \\
\s{\CMEC} = 0 & 0 & 2 & 4 & 5 & 4 & 6 & 8 & 8 & 10
\end{array}
\end{equation*}
Note that for $\s{x}[1..9]$ and $\s{x}[1..10]$, there are actually
{\it two} borders that cover a maximum number of positions;
in each case the border of minimum length is identified in $\s{\MEC}$.

\begin{figure}[h!]
{\leftskip=0.80in\obeylines\sfcode`;=3000
\bproc Compute\_MEC$(\s{\pi};\ \s{\MEC},\s{\CMEC})$
$n \la |\s{\pi}|$
Compute\_MNC$(n,\s{\pi};\ \B,\s{\gamma},\s{\MNC})$
$\s{\MEC} \la 0^n;\ \s{\CMEC} \la 0^n;\ \s{\PR} \la 1^{\B}$
\bfor $q \la 1$ \bto $\B$ \bdo $\CPR[q] \la q$
\bfor $i \la 2$ \bto $n$ \bdo
\qq $q \la \s{\pi}[i]$
\com{$\s{x}[i..i\+ q\m 1] = \s{x}[1..q]$.}
\qq \bwhile $q > 0$ \bdo
\com{$\s{x}[1..q']$ is the longest prefix of $\s{x}[1..q]$ without a cover.}
\qq\qq $q' \la \s{\MNC}[q]$
\com{$\s{x}[1..q']$ also occurs at $i$: update $\s{\CPR}[q']$ \& $\s{\PR}[q']$.}
\qq\qq \bif $i\m \s{\PR}[q'] < q'$ \bthen
\qq\qq\qq $\s{\CPR}[q'] \la \s{\CPR}[q']\+ i\m \s{\PR}[q']$
\qq\qq \belse
\qq\qq\qq $\s{\CPR}[q'] \la \s{\CPR}[q']\+ q'$
\qq\qq $\s{\PR}[q'] \la i$
\com{Update \s{\CMEC} \& \s{\MEC} accordingly for interval $i..i\+ q'\m 1$.}
\qq\qq \bif $\s{\CPR}[q'] \ge \s{\CMEC}[i\+ q'\m 1]$ \bthen
\qq\qq\qq $\s{\MEC}[i\+ q'\m 1] \la q'$
\qq\qq\qq \bif $\s{\CPR}[q'] > \s{\CMEC}[i\+ q'\m 1]$ \bthen
\qq\qq\qq\qq $\s{\CMEC}[i\+ q'\m 1] \la \s{\CPR}[q']$
\qq\qq $q \la q'\m 1$
}
\caption{Computing $\s{\MEC}$ amd $\s{\CMEC}$ from the prefix array $\s{\pi}$.}
\label{fig-MEC}
\end{figure}

The algorithm Compute\_MEC is shown in Figure~\ref{fig-MEC}.
In the first stage, $\B$ and $\s{\MNC}$ are computed
and the arrays \s{\CMEC}, \s{\PR} and \s{\CPR} are initialized.
Then every position $i > 1$ such that $q = \s{\gamma}[i] > 0$ is considered.
Using \s{\MNC}, the longest prefix $\s{\Q'} = \s{x}[1..q']$ of $\s{x}[1..q]$
that does {\it not} have a cover is identified;
for prefixes of $\s{x}[1..q]$ that do have a cover,
the appropriate \s{\PR} and \s{\CPR} values have already been updated.
There are two main steps in the processing of \s{\Q'}:
\begin{itemize}
\item[$\bullet$]
Since $i$ has now become the rightmost occurrence of \s{\Q'} in $\s{x}[1..i]$,
we must set $\s{\PR}[q'] \la i$ and increment the corresponding number $\s{\CPR}[q']$
of positions covered.
\item[$\bullet$]
If the number $\s{\CPR}[q']$ of positions covered by occurrences of \s{\Q'}
exceeds $\s{\CMEC}[i\+ q\- 1]$,
then \s{\CMEC} and \s{\MEC} must be updated accordingly.
\end{itemize}
These steps are repeated recursively for the longest proper prefix of $\s{\Q'}$
that does not have a cover.

\section{Correctness \& Complexity of Compute\_MEC}
\label{sect-analysis}
We begin by proving the correctness of Compute\_MEC,
which depends on the prior computation of $\s{\pi} = \s{\pi}_{\s{x}}$
\cite{BKS13}.
Consider first procedure Compute\_MNC,
where $\B$ is computed, followed by the cover array $\s{\gamma}[1..\B]$.
Then for every $q \in 1..\B$, $\s{\MNC}[q] \la q$
whenever there is no cover of $\s{x}[1..q]$,
with $\s{\MNC}[q]\ \la \s{\MNC}[q\- 1]$ otherwise,
an easy and straightforward calculation.

Compute\_MEC then independently considers positions $i = 2,3,\ldots,n$
for which $\s{\pi[i]} > 0$; that is,
such that a border of \s{x} of length $q = \s{\pi}[i]$ begins at $i$.
The internal \bwhile loop then processes in decreasing order of length
the prefixes $\s{\Q'} = \s{x}[1..q']$
of $\s{x}[1..q]$ that have no cover --- and that therefore,
by Observation~\ref{obs-mnc},
can possibly be minimum enhanced covers of $\s{x}[1..i\+ q'\m 1]$.
Thus, for every $i \in 2..n$, all such borders
$\s{x}[1..q] = \s{x}[i..i\+ q\- 1]$ are considered and,
for each one, all such prefixes \s{\Q'}.
For each $q'$:
\begin{itemize}
\item[$\bullet$]
the number $\CPR[q']$ of positions covered by $\s{\Q'}$ is updated,
as well as the position $\PR[q'] = i$ of rightmost occurrence of \s{\Q'};
\item[$\bullet$]
$\MEC[i\+ q'\- 1]$ and $\CMEC[i\+ q'\- 1]$ are
updated accordingly for sufficiently large $\CPR[q']$.
\end{itemize}

We claim therefore that
\begin{thrm}\label{thrm-time}
For a given string \s{x},
Compute\_MEC correctly computes the minimum enhanced cover array $\s{\MEC}_{\s{x}}$
and the number $\s{\CMEC}_{\s{x}}$ of positions covered by it,
based solely on the prefix array $\s{\pi}_{\s{x}}$.
\end{thrm}

We have seen that
in aggregate Compute\_MEC processes a subset of the nonempty borders
of every prefix $\s{x}[1..i]$,
devoting $\mathcal{O}(1)$ time to each one.
As we have seen,
each border \s{\Q'} in each such subset is constrained to have no cover.
We say that a string \s{v} is \itbf{strongly periodic} if
it has a border \s{u} such that $|\s{u}| \ge |\s{v}|/2$;
otherwise \s{v} is said to be \itbf{weakly periodic}.
Observe that the borders \s{\Q'} must all be weakly periodic;
if not, then they would have a cover \s{u} with $|\s{u}| \ge |\s{v}|/2$.
In \cite{FIKPPST13} the following result is proved:
\begin{lemm}
\label{lemm-weak}
There are at most $\log_2 n$ weakly periodic borders of
a string of length $n$.
\end{lemm}
It follows then that for each $i \in 2..n$,
there are at most $\log_2 i$ borders considered,
thus overall $\mathcal{O}(n\log n)$ time.

The space requirement of Compute\_MEC,
apart from the $\s{\pi}$, \s{\MEC} and \s{\CMEC} arrays,
each of length $n$,
consists of three integer arrays
(\s{\MNC} (overwriting \s{\gamma}), \s{\PR}, \s{\CPR}),
each of length $\B < n$.
Thus
\begin{thrm}
\label{thrm-space}
In the worst case, Compute\_MEC computes
\s{\MEC} and \s{\CMEC} from \s{\pi} using
\begin{itemize}
\item[$(a)$]
$\mathcal{O}(n\log n)$ time;
\item[$(b)$]
three additional arrays $1..\B$ of integers $1..n$,
thus $\Theta(\B\log n)$ bits of space.
\end{itemize}
\end{thrm}

Now consider the expected (average) case behaviour
of Compute\_MEC. This depends critically on the expected length of the maximum border
of $\s{x}[1..n]$; that is, the expected value of $\B$.
We show in the Appendix that for a given alphabet size,
$\B$ approaches a limit as $n$ goes to infinity.
The limit is approximately 1.64 for binary alphabets, 0.69 for ternary alphabets,
and monotone decreasing in alphabet size.
Thus
\begin{thrm}
\label{thrm-exp}
In the average case, Compute\_MEC requires $\mathcal{O}(n)$ time
and $\Theta(\log n)$ additional bits of space.
\end{thrm}

\section{Comparing Border-Based and Prefix-Based Algorithms}
\label{sect-compare}
As has been mentioned above, in order to compute $\MEC_{\s{x}}$, the authors of
\cite{FIKPPST13} made use of the \itbf{border array}. On the other hand
Compute\_MEC is based on the \itbf{prefix table}. We have already highlighted
the advantage Compute\_MEC has because of the use of a \itbf{prefix table} in
lieu of a \itbf{border array} especially in the context of indeterminate
strings. Additionally, the simplicity and low space usage of Compute\_MEC
encourage us to compare its practical performance with the algorithm of
\cite{FIKPPST13}. To this end this can be seen as a comparison between a
border-based algorithm (i.e., the algorithm of \cite{FIKPPST13}) for computing
$\MEC_{\s{x}}$ and a prefix-based algorithm (i.e, Compute\_MEC of the current
paper) for doing the same. In what follows we will refer to the former algorithm
as ECB and the latter as ECP.

\begin{figure}[t!]
  \centering
  \includegraphics*[scale=0.6]{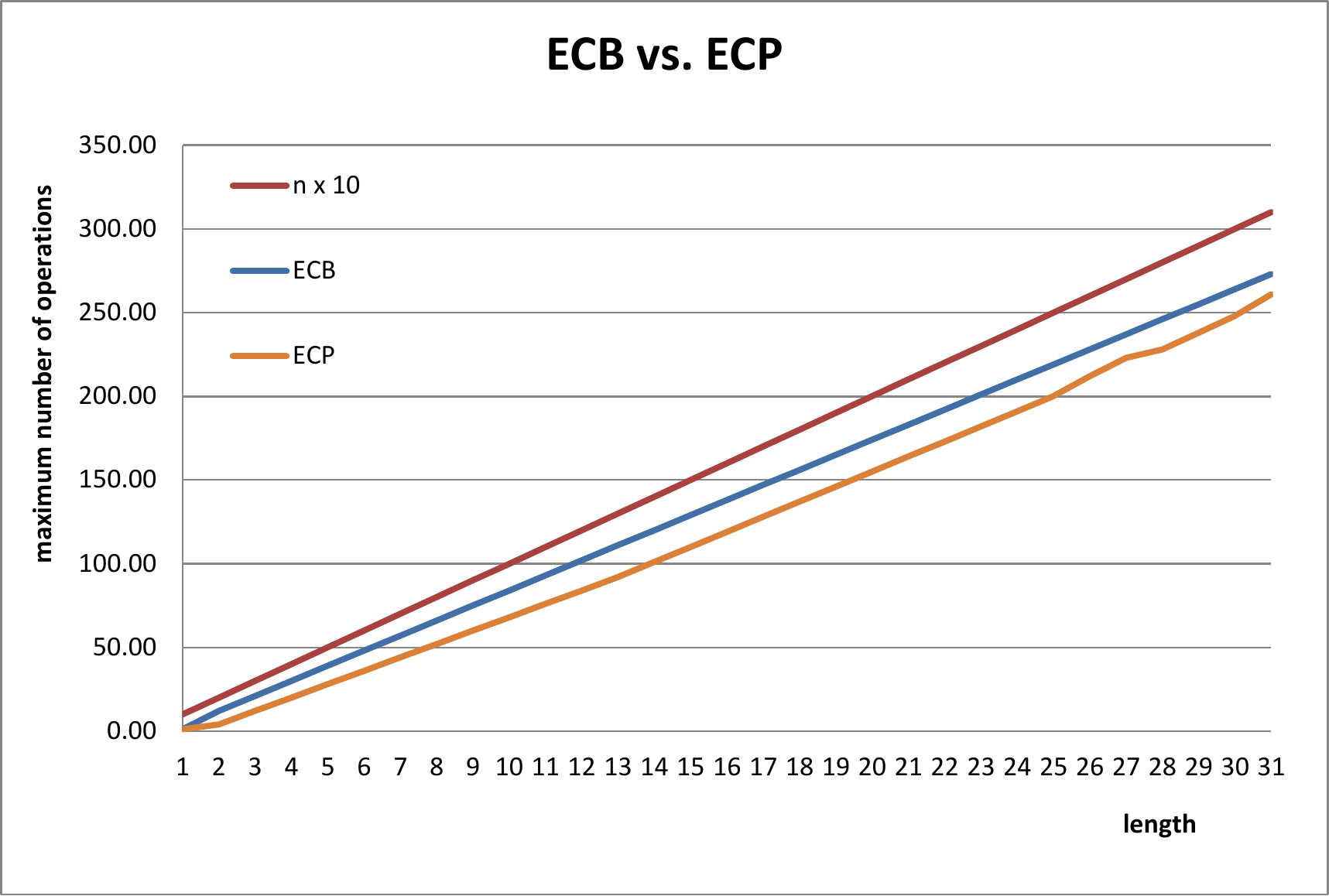} 
  \caption{
   The maximum number of operations performed by the Border-Based (ECB)
   \cite{FIKPPST13} and Prefix-Based (ECP) algorithm (i.e., Compute\_MEC) to
   compute the Minimum Enhanced Cover array, for all strings on the binary alphabet.}
  \label{fig-exp1}
\end{figure}

\begin{figure}[t!] 
  \centering
  \includegraphics*[scale=0.6]{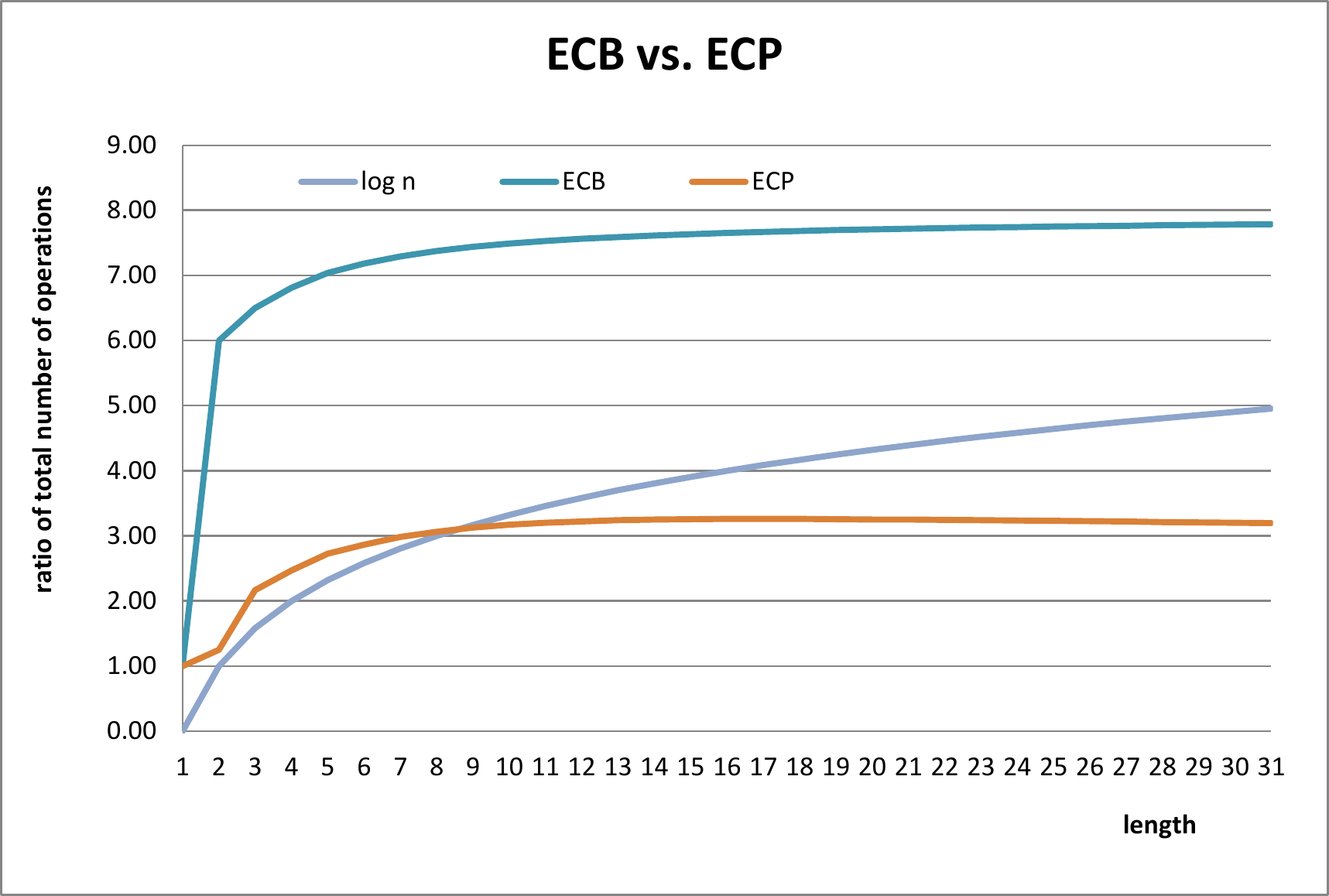}
  \caption{
   Ratio of the total number of operations performed by the Border-Based (ECB)
   \cite{FIKPPST13} and Prefix-Based (ECP) algorithm to the
   length $n$ of string, for all strings on the binary alphabet.}
  \label{fig-exp2}
\end{figure}

We have implemented ECP (i.e., Compute\_MEC) in C\# using Visual 
Studio 2010. We got the implementation of ECB from the authors of
\cite{FIKPPST13}. However, ECB was implemented in C. To ensure a level playing
ground, we re-implemented ECB in C\# following their implementation. Then we
have run both the algorithms on all binary strings of lengths 2 to 30. The
experiments have been carried out on a Windows Server 2008 R2 64-bit Operating
System, with Intel(R) Core(TM) i7 2600 processor @ 3.40GHz having an installed
memory (RAM) of 8.00 GB. 
The results are illustrated in Figure \ref{fig-exp1} and \ref{fig-exp2}, 
where the maximum number of operations 
carried out by each algorithm is reported in Figure \ref{fig-exp1}. Figure
\ref{fig-exp2} shows the ratio of the total number of operations performed by
the Border-Based (ECB) \cite{FIKPPST13} and Prefix-Based (ECP) algorithm to the
length $n$ of string, for all strings on the binary alphabet.
As is evident from
Figure~\ref{fig-exp1} and \ref{fig-exp2} , ECP outperforms ECB and in fact it
does show a linear behaviour verifying the claim in Theorem \ref{thrm-exp} above.

%
%
\section{Extensions}
\label{sect-extend}
In Sections~\ref{sect-method} and \ref{sect-analysis}
we describe an algorithm to compute the minimum
enhanced cover array $\MEC_{\s{x}}$ of a given string \s{x},
based only on the prefix array $\s{\pi}_{\s{x}}$.
As noted in the Introduction,
since the prefix array can be computed also for indeterminate strings \cite{SW08},
this immediately raises the possibility of extending
the $\MEC$ calculation to indeterminate strings.

In \cite{ARS14a} two definitions of ``cover'' for an indeterminate string
are proposed:
a \itbf{sliding cover} where adjacent or overlapping covering substrings of \s{x} must match,
and a \itbf{rooted cover}
where each covering substring is constrained only to match a prefix of \s{x}.
The nontransitivity of matching (see Section~\ref{sect-intro})
inhibits implementation of a sliding cover,
but \cite{ARS14a} shows how to compute all the rooted covers
of indeterminate \s{x} from its prefix array
in $\mathcal{O}(n^2)$ worst case time,
$\Theta(n)$ in the average case.
Thus it becomes possible to execute Compute\_MNC for rooted covers,
simply by replacing the function call to Compute\_PCR by a function
call to PCInd of \cite{ARS14a};
that is, to compute the rooted cover array $\s{\gamma_{_R}}[1..\B]$,
hence $\s{\MNC}[1..\B]$ and thus $\s{\MEC}_{\s{x}}$,
all for indeterminate strings.
Let us call this new algorithm Compute\_MEC\_Ind.
We recall now a lemma from \cite{BRS09} stating that the
expected number of borders in an indeterminate string is bounded above by a constant, approximately 29.
Therefore, also for indeterminate strings, $\B$ can be treated as a constant,
and we have the following remarkable result:
\begin{thrm}
\label{thrm-expind}
In the average case, Compute\_MEC\_Ind requires $\mathcal{O}(n)$ time
and $\Theta(\log n)$ additional bits of space.
\end{thrm}

We note further that the prefix array can be efficiently computed in
a compressed form \cite{SW08}, taking advantage of the fact
that for $i \in 1..n$,
$\s{\pi}[i] \ne 0$ if and only if $\s{x}[i] = \s{x}[1]$.
Thus we can use two arrays $\POS$ and $\LEN$
to store nonzero positions in \s{\pi} and
the values at those positions, respectively,
thus saving much space in cases that arise in practice.
We have designed a $\POS/\LEN$ version of Compute\_MEC
that space restrictions do not allow us to describe here.

Finally, \cite{FIKPPST13} describes extensions of the minimum enhanced cover array
calculation, as follows:
\begin{itemize}
\item[$\bullet$]
computation of the enhanced left-cover array of \s{x};
\item[$\bullet$]
computation of the enhanced left-seed array of \s{x}.
\end{itemize}
Our prefix array approach yields efficient algorithms
for these problems also,
that may similarly be extended
to rooted covers of indeterminate strings.

\def\AJC{Australasian J.\ Combinatorics\ }
\def\AWOCA{Australasian Workshop on Combinatorial Algs.}
\def\CPM{Annual Symp.\ Combinatorial Pattern Matching}
\def\COCOON{Annual International Computing \& Combinatorics Conference}
\def\FOCS{IEEE Symp.\ Found.\ Computer Science}
\def\AESA{Annual European Symp.\ on Algs.}
\def\LATA{Internat.\ Conf.\ on Language \& Automata Theory \& Applications}
\def\IWOCA{Internat.\ Workshop on Combinatorial Algs.}
\def\AWOCA{Australasian Workshop on Combinatorial Algs.}
\def\STACS{Internat.\ Symp.\ Theoretical Aspects of Computer Science}
\def\ICALP{Internat.\ Colloq.\ Automata, Languages \& Programming}
\def\IJFCS{Internat.\ J.\ Foundations of Computer Science\ }
\def\ISAAC{Internat.\ Symp.\ Algs.\ \& Computation}
\def\SPIRE{String Processing \& Inform.\ Retrieval Symp.}
\def\SWAT{Scandinavian Workshop on Alg.\ Theory}
\def\PSC{Prague Stringology Conf.}
\def\WALCOM{International Workshop on Algorithms \& Computation}
\def\ALG{Algorithmica\ }
\def\CSUR{ACM Computing Surveys\ }
\def\FI{Fundamenta Informaticae\ }
\def\IPL{Inform.\ Process.\ Lett.\ }
\def\IS{Inform.\ Sciences\ }
\def\JACM{J.\ Assoc.\ Comput.\ Mach.\ }
\def\CACM{Commun.\ Assoc.\ Comput.\ Mach.\ }
\def\MCS{Math.\ in Computer Science\ }
\def\NJC{Nordic J.\ Comput.\ }
\def\SICOMP{SIAM J.\ Computing\ }
\def\SIDMA{SIAM J.\ Discrete Math.\ }
\def\JCB{J.\ Computational Biology\ }
\def\JA{J.\ Algorithms\ }
\def\JCMCC{J.\ Combinatorial Maths.\ \& Combinatorial Comput.\ }
\def\JDA{J.\ Discrete Algorithms\ }
\def\JALC{J.\ Automata, Languages \& Combinatorics\ }
\def\SODA{ACM-SIAM Symp.\ Discrete Algs.\ }
\def\SPE{Software, Practice \& Experience\ }
\def\TCJ{The Computer Journal\ }
\def\TCS{Theoret.\ Comput.\ Sci.\ }

\section*{APPENDIX}

We write $|\s x|$ for the length of string $\s x$.  Here we show that the expected length
of the longest border of a string \s{x} approaches a limit as $|\s x |$ tends to infinity, the limit depending on the alphabet size. For a binary alphabet it is approximately 1.64.  We use the following notation.
$\sigma=|\Sigma|$ is the alphabet size, $B(w)$ is length of longest border of string $w$ and
$B_k(w)$ is length of longest border of string $w$ which has length at most $k$  (ie, ignoring any borders longer than $k$). Thus if $\itbf{x}= babaabababbabaabab$ then $B(\itbf{x})=8$ since $\itbf{x}$ has longest border $babaabab$ and $B_4(\itbf{x})=3$ since the longest border of $\itbf{x}$ which has length at most 4 is $aba$.
$W_n$ is the set of all strings of length $n$ on an alphabet of size $\sigma$. Since $W_0$ contains only the empty string we have $|W_0|=1$.\\

\begin{lemm}\label{l1} The number of strings of length $n$ on an alphabet of
size $\sigma$ which have a border of length $k$
(not necessarily the longest border) is $\sigma^{n-k}$.
\begin{proof} A string with border of length $k$ is periodic with period
$n-k$ and so is determined by its  length $n-k$ prefix. This prefix can be
chosen in $\sigma^{n-k}$ ways. \hfill $\Box$
\end{proof}
\end{lemm}
We also need the following formula (which can be obtained using a computer algebra system).
\begin{lemm}\label{l2}
$\sum_{i=a}^b m\sigma^m={\frac {{\sigma}^{b+1} \left( \sigma\, \left( b+1 \right) -\sigma-b-1
 \right) }{ \left( \sigma-1 \right) ^{2}}}-{\frac {{\sigma}^{a}
 \left( a\sigma-a-\sigma \right) }{ \left( \sigma-1 \right) ^{2}}}$.
\end{lemm}

Clearly $|W_n|=\sigma^n$.  The expected size of the longest border of a string of length $n$ on an alphabet of size $\sigma$ is therefore
\begin{equation}\label{e0}
\overline{B}(n)=\frac{1}{\sigma^n}\sum_{w \in W_n} B(w). 
\end{equation} Similarly, the expected size of the longest border not exceeding $k$ is
\begin{equation}\label{e0.5}
\overline{B}_k(n)=\frac{1}{\sigma^n}\sum_{w \in W_n} B_k(w).
\end{equation}
 Clearly $B(w) \ge B_k(w)$ so
\begin{equation} \label{e1}
\overline{B}(n) \ge \overline{B}_k(n).
\end{equation} Note that if $n \ge 2k$ then $W_n=\{uvw:u \in W_k,x \in W_{n-2k},v \in W_k\}$ and so
\begin{equation} \label{e2}
\overline{B}_k(n) =\frac{1}{\sigma^n}\sum_{u \in W_k}\sum_{x\in W_{n-2k}}\sum_{v \in W_k}B_k(uxv).
\end{equation} Now $B_k(uxv)=B_k(uv)$ so if $n \ge 2k$,
\begin{eqnarray} \label{e3}
\overline{B}_k(n) &=&\frac{1}{\sigma^n}\sum_{u \in W_k}\sum_{v \in W_k}B_k(uv)\sum_{x\in W_{n-2k}}1 \\
\nonumber &=&\frac{\sigma^{n-2k}}{\sigma^n}\sum_{u \in W_k}\sum_{v \in W_k}B_k(uv)\\
\nonumber &=& \frac{1}{\sigma^{2k}}\sum_{w \in W_{2k}}B_k(w)\\
\nonumber &=&\overline{B}_k(2k).
\end{eqnarray} With (\ref{e1}) we then have, for $n \ge 2k$,
\begin{equation} \label{e4}
\overline{B}(n) \ge \overline{B}_k(2k).
\end{equation}

Now any border that is counted in the right hand side of (\ref{e0}) but not
counted on the right hand side of (\ref{e0.5}) has length at least $k+1$.
The sum of the lengths of such borders is, by Lemma \ref{l1},
$$\sum_{m=k+1}^nm\sigma^{n-m}.$$ So, by Lemma \ref{l2} and (\ref{e3}),
\begin{eqnarray}\label{upper bound}
\overline{B}(n) &\le & \frac{1}{\sigma^n}(\sum_{w \in W_n} 
                            B_k(w)+\sum_{m=k+1}^nm\sigma^{n-m})\\
\nonumber &=& \overline{B}_k(n)+\frac{1}{\sigma^n}({\frac {{\sigma}^{n-k+1}k+{\sigma}^{n-k+1}-{\sigma}^{n-k}k-\sigma\,n-
\sigma+n}{ \left( \sigma-1 \right) ^{2}}}
)\\
\nonumber &<& \overline{B}_k(n)+{\frac {{\sigma}^{-k+1}k+{\sigma}^{-k+1}-{\sigma}^{-k}k-
\sigma}{ \left( \sigma-1 \right) ^{2}}}\\
\nonumber &=&\overline{B}_k(2k)+O(k\sigma^{-k}).
\end{eqnarray}

Thus for all $n\ge 2k$ $$\overline{B}_k(2k) \le \overline{B}(n) \le
\overline{B}_k(2k)+O(\sigma^{-k})$$ so they're contained in an arbitrarily
small interval. Call this interval $I_k$ and define $J_1=I_1$ and for $i
\ge 2$ $J_i=I_i \cap J_{i-1}$. Then $J_1,J_2,\dots$ is a sequence of nested
intervals whose lengths have limit 0. By the Nested Intervals Theorem this
means the limit of $\overline{B}_n$ exists.

Using (\ref{e1})  and (\ref{upper bound}) with $k=11 $ we find that $lim_{n
\rightarrow \infty} \overline{B}(n)$ lies in the interval $(1.6356,1.6420)$
for binary alphabets.  For ternary alphabets using $k=6$ the limit lies in
$(0.6811,0.6864)$.

\end{document}